\def\be{\begin{equation}}
\def\ee{\end{equation}}

\documentstyle[eqsecnum,aps,preprint]{revtex}
\begin{document}
\draft
\title{Structure of the  gluon propagator at finite temperature}
\author{H. Arthur Weldon}
\address{Department of Physics, West Virginia University, Morgantown WV
26506-6315}
\date{December 16, 1996}
\maketitle

\begin{abstract}
The thermal self-energy of gluons generally depends on four Lorentz-invariant
functions. Only two of these occur in the hard thermal loop approximation of
Braaten and Pisarski because of the abelian Ward identity 
$K_{\mu}\Pi^{\mu\nu}_{\rm htl}=0$.
However, for the exact self-energy  $K_{\mu}\Pi^{\mu\nu}\neq 0$. In linear gauges the
Slavnov-Taylor identity is shown to require a non-linear relation among three of the
Lorentz-invariant self-energy function:  
$(\Pi_{C})^{2}=(K^{2}-\Pi_{L})\Pi_{D}$. This reduces the exact gluon propagator
to a simple form containing only two types of poles: one that determines the behavior of 
transverse electric and magnetic gluons and one that controls the longitudinally polarized electric
gluons.

\end{abstract}

\narrowtext

\section{Introduction}

\subsection{Background}

The constraint imposed by local gauge invariance  on the various Green
functions of gauge theories are contained in  the Slavnov-Taylor
identities \cite{Slav,JC}. These identities were originally formulated in covariant gauge, which
has a gauge-fixing term  ${\cal L}_{\rm gf}=-(\partial_{\mu}A^{\mu})^{2}/2\xi$.
In  covariant  gauge the  identity satisfied by the photon
propagator of QED is
 \be K_{\mu}D^{\prime\mu\nu}_{\rm qed}(K)=-\xi {K^{\nu}\over
K^{2}},\ee 
whereas  the identity satisfied by the gluon propagator of QCD is slightly weaker:
\be K_{\mu}K_{\nu}D^{\prime\mu\nu}(K)=-\xi.\label{1.2} \ee
At zero temperature, the abelian identity (1.1) has the same consequence as
the nonabelian version (1.2)  for  the respective self-energy tensors. In both
cases $\Pi^{\mu\nu}$   must be a
 a linear combination of the available tensors $g^{\mu\nu}$ and $K^{\mu}K^{\nu}$.
When the propagator is expressed in terms of the self-energy, application of (1.1)
yields $K_{\mu}\Pi^{\mu\nu}_{\rm qed}=0$ for the abelian case and application of
(1.2) yields $K_{\mu}\Pi^{\mu\nu}=0$ for QCD.

At non-zero temperature the Slavnov-Taylor identities are unchanged, as
shown in  Appendix C. However the self-energy tensor can depend on the
four-velocity $u^{\mu}$ of the plasma. Consequently in both QED and QCD the
self-energy tensor can be  a linear combination of four possible tensors:
$g^{\mu\nu}, K^{\mu}K^{\nu}, u^{\mu}u^{\nu},$ and
$K^{\mu}u^{\nu}+u^{\mu}K^{\nu}$.  It is convenient to choose particular linear
combinations of these as the standard basis set \cite{GPY,KK,rev} denoted 
$A^{\mu\nu}, B^{\mu\nu}, C^{\mu\nu}, D^{\mu\nu}$ and defined in Appendix A. 
By parametrizing the tensor structure of the
self-energy one can then calculate the form of the exact propagator and apply the
Slavnov-Taylor indentity. For QED the result remains 
$K_{\mu}\Pi^{\mu\nu}_{\rm qed}=0,$
since the electromagnetic current operator
satisfies $\partial_{\mu}j^{\mu}=0$ even at non-zero temperature. Only two of
the four basis tensors satisfy this. Therefore
\be
\Pi^{\mu\nu}_{\rm qed}=\Pi^{\rm qed}_{T}A^{\mu\nu}+\Pi^{\rm
qed}_{L}B^{\mu\nu},\ee
where the subscripts $T$ and $L$ denote transverse and longitudinal with respect
to the spatial component $\vec{k}$ of the wave-vector.

For QCD Braaten and Pisarski \cite{BP1,BP2} discovered a consistent 
 high-temperature approximation which  behaves in the same fashion as quantum
electrodynamics. 
In this hard thermal loop approximation the quarks are massless and only the order
$T^{2}$ part of  one-loop diagrams are retained. The order $T^{2}$ part of the
one-loop gluon self-energy is  gauge invariant and  satisfies  
$K_{\mu}\Pi^{\mu\nu}_{\rm htl}=0.$
In the standard basis  
$\Pi^{\mu\nu}_{\rm htl}=\Pi_{T}^{\rm htl}
A^{\mu\nu}+\Pi_{L}^{\rm htl}B^{\mu\nu}$,
and the functions $\Pi_{T}, \Pi_{L}$ are gauge invariant. 
This  was  discovered by
explicit calculation in covariant gauges \cite{KKW} and was  shown by Braaten and Pisarski to be a
consequence of the fact that the hard thermal loops with any number of
external gluons all satisfy abelian Ward identities \cite{BP1,BP2,F&T,act}.  
In hard thermal loop approximation the gluon propagator has the same structure
as the photon thermal propagator.

If one goes beyond the hard thermal loop approximation to QCD 
these simplifications do not apply. Inclusion of sub-dominant powers of $T$, 
quark masses, or higher loop effects all spoil the abelian features.
For example, even at one-loop order  with
massless quarks,  the gluon self-energy  has a sub-dominant term linear
in $T$ that yields $K_{\mu}\Pi^{\mu\nu}\neq 0$ \cite{KKM}.  
Thus, generally
\be
K_{\mu}\Pi^{\mu\nu}\neq 0.\ee
The QCD self-energy therefore depends on all four  basis tensors:
 \be
\Pi^{\mu\nu}=\Pi_{T}A^{\mu\nu}+\Pi_{L}B^{\mu\nu}+\Pi_{C}C^{\mu\nu}
+\Pi_{D}D^{\mu\nu},\label{pi}\ee
where  $K_{\mu}C^{\mu\nu}\neq 0$ and
$K_{\mu}D^{\mu\nu}\neq 0$. In the exact self-energy the functions $\Pi_{T},\Pi_{L},
\Pi_{C}, \Pi_{D}$ are all gauge-dependent. 

\subsection{Summary of Results}

The purpose of this paper is to investigate the structure of the full gluon
propagator at finite temperature that results from the self-energy (\ref{pi}).
In a general linear gauge the nonabelian Slavnov-Taylor identity (\ref{1.2})
generalizes to
\be
F_{\mu}F_{\nu}D^{\prime\mu\nu}=-\xi,\label{STI}\ee
where  $F^{\mu}$ is the gauge-fixing vector.
The obvious way of implementing this is to
insert the free propagator with gauge-fixing vector $F^{\mu}$ and the self-energy (\ref{pi}) into the
Schwinger-Dyson equation for the full propagator $D^{\prime\mu\nu}$. After some tedious algebra is
performed the Slavnov-Taylor constraint (\ref{STI})  can be be applied.  

A more direct procedure will be followed here. 
The quadratic part of the gluon action is the sum of the
kinetic energy, the self-energy, and the gauge-fixing term.
The reduced action without the gauge-fixing term will be denoted by a check:
 \be
\check{\Gamma}^{\mu\nu}
=-K^{2}g^{\mu\nu}+K^{\mu}K^{\nu}+\Pi^{\mu\nu}.\ee
Sec III will show that the Slavnov-Taylor identity (\ref{STI}) can be expressed as the pair of
constraints:
\be
\check{\Gamma}^{\mu\nu}G_{\nu}=0\hskip1cm 
F^{\nu}G_{\nu}=-\xi,\label{pair}\ee
which determine the vector $G^{\nu}$.
The consequences of this are simple.
Viewed as a  matrix,  $\check{\Gamma}^{\mu}_{\;\;\nu}$ must have four eigenvalues.
Three eigenvalues are determined dynamically by the equations of motion. The
fourth eigenvalue must be zero because of local gauge
invariance as expressed in (\ref{pair}).  A zero eigenvalue is only possible if 
the determinant of  $\check{\Gamma}^{\mu}_{\;\;\nu}$ vanishes. This immediately gives a non-linear
relation on the gluon self-energy.

The condition Det$(\check{\Gamma}^{\mu}_{\;\;\nu})=0$ applies with any linear gauge-fixing conditon.
If we specialize to gauges that preserve
the  rotational invariance of the plasma rest frame, then
the gauge-fixing vector $F^{\mu}$ should be a linear combination of $K^{\mu}$ and the plasma velocity
$u^{\mu}$. In all such cases the self-energy has the tensor structure (\ref{pi}).
The requirement Det$(\check{\Gamma}^{\mu}_{\;\;\nu})=0$ leads to
  \be
(\Pi_{C})^{2}=(K^{2}-\Pi_{L})\Pi_{D},\label{p2}\ee
which must hold  for all $K^{\mu}$. This is quite remarkable since the functions
$\Pi_{L}, \Pi_{C}, \Pi_{D}$ are all gauge-dependent.
In a perturbative expansion, $\Pi_{C}$ is known to begin at order $g^{2}$
\cite{KKM} and
consequently  $\Pi_{D}$ must begin at order $g^{4}$. Results equivalent to this
were obtained
 by Gross, Pisarski, and Yaffe \cite{GPY} and by Kajantie and Kapusta \cite{KK}in
covariant gauge and in general linear gauges by Kobes, Kunstatter, and Mak
\cite{KKM}.

Because of the simplification (\ref{p2}) it is rather easy to compute the 
full gluon propagator from the Schwinger-Dyson equation. The full propagator is constructed of four
tensors but it does not have four poles. Instead there are only two:
one pole  associated with the two transverse polarizations contained in $A^{\mu\nu}$ that determines
the behavior of  transverse electric and magnetic gluons; and one pole associated with
a particular combination of $B^{\mu\nu}, C^{\mu\nu}, D^{\mu\nu}$ that controls the longitudinal
electric gluons.

The entire development is done in the real-time formulation of thermal QCD.
The full gluon propagator has a $2\times 2$ matrix structure
$D^{\prime\mu\nu}_{ab}$, where $a,b=1,2$. The formulas displayed 
throughout this paper are for the diagonal entries of the $2\times 2$ matrix
structure. The diagonal entries satisfy Schwinger-Dyson equations that involve
products of functions rather than products of matrices. For example, if $k_{0}$ is
given a small positive imaginary part, i.e. $k_{0}+i\epsilon$, then all formulas
in the paper apply to retarded propagators and self-energies. If
instead, $k_{0}$ is given an imaginary part with alternating sign, i.e.
$k_{0}+i\epsilon k_{0}$, then all formulas apply to  propagators and self-energies
that satisfy Feynman boundary conditions. Appendix C gives further details.

In Sec II the exact form of the thermal gluon propagator is displayed for for three popular gauges:
covariant, Coulomb, and temporal. The proof of these results is delayed until  Sec III, which derives
the general result (\ref{exact}) for the thermal gluon propagator  for any linear gauge-fixing
condition that preserves the rotational invariance of the plasma rest frame. Sec IV discusses the
gauge invariance of the location of the transverse and longitudinal  poles 
and shows that the two types of poles coincide at $\vec{k}=0$.  Appendix A gives
explicit details about the four basis tensors. Appendix B derives the form of the gluon propagator in
the hard thermal loop approximation for a general linear gauge. 
Appendix C derives the Slavnov-Taylor identity for
general linear gauges at finite temperature.

\section{Gluon propagator in three gauges}

The final result for the full gluon propagator in a general linear gauge is given in (\ref{exact}). 
Before deriving the general form it may be helpful to display  the final result in
familiar gauges, viz.  covariant, Coulomb, and temporal. To give these results,
some notation is necessary. 
 Instead of the four-vectors $K^{\mu}$  and $u^{\mu}$ it is
simpler to use $K^{\mu}$ and $\tilde{K}^{\mu}$ where
\be
\tilde{K}^{\mu}=(K^{\mu} K\cdot u-u^{\mu}K^{2})/k,\label{tilde}\ee 
and $ k^{2}\equiv (K\cdot u)^{2}-K^{2}\ge 0$. This
 vector satisfies 
\be
 K\cdot\tilde{K}=0\hskip1cm\tilde{K}^{2}=-K^{2}.\ee 
It is often convenient to use the plasma rest frame $u^{\mu}=(1,0,0,0)$ in which case 
$\tilde{K}^{\mu}=(k,\hat{k}k_{0})$. In this frame the transverse ternsor $A^{\mu\nu}$ has
the following value
\be
A^{\mu\nu}\Big|_{\rm rest}=\left(\matrix{ 0&& 0\cr 0 &&
-\delta^{mn}+\hat{k}^{m}\hat{k}^{n}\cr}\right).\ee
A second notation requires solving (\ref{p2}) for $\Pi_{C}$:
\be
\Pi_{C}=\sigma\sqrt{(K^{2}-\Pi_{L})\Pi_{D}},\label{root}\ee
where $\sigma=\pm1$. In terms of this, it will be useful to employ
 the four-vector 
\be H^{\mu}=K^{\mu}\sqrt{K^{2}-\Pi_{L}}+\sigma\tilde{K}^{\mu}\sqrt{\Pi_{D}}
\label{W},\ee
which has components $H^{\mu}
=(h_{0}, \hat{k}\,h)$. 

\subsection{Covariant gauges}

In covariant gauges the free
propagator is
\be
 D^{\mu\nu}_{\rm free}=\Bigl[-g^{\mu\nu}+(1-\xi){K^{\mu}K^{\nu}\over
K^{2}}\Bigr]{1\over K^{2}}.\ee 
or equivalently
\be
D^{\mu\nu}_{\rm free}=-{A^{\mu\nu}\over
K^{2}}+{\tilde{K}^{\mu}\tilde{K}^{\nu}\over (K^{2})^{2}}-\xi{K^{\mu}K^{\nu}\over (K^{2})^{2}}.\ee 
In this gauge the exact propagator (\ref{exact}) reduces  to
\be D^{\prime\mu\nu}={-A^{\mu\nu}\over K^{2}-\Pi_{T}}
+{\tilde{K}^{\mu}\tilde{K}^{\nu}\over K^{2}(K^{2}-\Pi_{L})}
-{\xi H^{\mu}H^{\nu}\over (K^{2})^{2}(K^{2}-\Pi_{L})}.\label{cov}\ee
 At $K^{2}=\Pi_{T}$ there is a pole in the transverse sector. 
At  $K^{2}=\Pi_{L}$  there is a pole which appears to have a complicated tensor structure. However, 
as $K^{2}\to\Pi_{L}$ the auxiliary vector $H^{\mu}\to \sigma\sqrt{\Pi_{D}}\tilde{K}^{\mu}$. 
Consequently at this pole
\be (K^{2}-\Pi_{L})D^{\prime\mu\nu}\Big|_{\rm pole}\to\bigl[1-\xi{\Pi_{D}\over K^{2}}\bigr]
{\tilde{K}^{\mu}\tilde{K}^{\nu}\over K^{2}},\ee
and the non-transverse function $\Pi_{D}$ only modifies the residue of the pole.

\subsection{Coulomb gauges}

A second example is provided by
 generalized Coulomb gauges with gauge-fixing term  ${\cal
L}_{\rm gf}=-({\bf \nabla}\cdot{\bf A})^{2}/2\xi$.  The 
free gluon propagator is 
\be D^{\mu\nu}_{\rm free}=-{A^{\mu\nu}\over K^{2}}
+{1\over k^{2}}\left(\matrix{ 1&&0\cr 0&&0\cr}\right)
-\xi{K^{\mu}K^{\nu}\over k^{4}}.\ee
In this case the $D^{00}$ propagator is instantaneous and only the transverse
potentials have time-dependence. The true Coulomb gauge $\nabla\cdot{\bf A}=0$
corresponds to $\xi=0$. 
 The full propagator that results from (\ref{exact}) is
 \be D^{\prime\mu\nu}={-A^{\mu\nu}\over K^{2}-\Pi_{T}}
+{K^{2}\over h^{2}}\left(\matrix{1 && 0\cr
0 && 0\cr}\right)-\xi{H^{\mu}H^{\nu}\over h^{2}k^{2}},\label{coul}\ee
where $h$ is the spatial component of $H^{\mu}$:
\be
h=k\sqrt{K^{2}-\Pi_{L}}+\sigma k_{0}\sqrt{\Pi_{D}}.
\label{wcomponents}\ee
As $h\to 0$, $H^{\mu}\to (h^{0}, 0)$ and $h^{0}\to -\sigma\sqrt{\Pi_{D}}K^{2}/k$. Consequently 
this pole also has a simple tensor structure: 
\be h^{2}\;D^{\prime\mu\nu}\Big|_{\rm pole}\to K^{2}[1-\xi\Pi_{D} {K^{2}\over k^{4}}]
\left(\matrix{1&&0\cr0&&0}\right).\ee

\subsection{Temporal gauges}

A third type of example is the generalized temporal gauge propagator.  The gauge-fixing term  is
${\cal L}_{\rm gf}=-(A^{0})^{2}/2\xi$, where $\xi$ has
 dimensions of length squared. The free propagator is 
\be 
D^{\mu\nu}_{\rm free}=-{A^{\mu\nu}\over K^{2}}
+{1\over k_{0}^{2}}\left(\matrix{0&&0\cr 0&&\hat{k}^{m}\hat{k}^{n}\cr}
\right)-\xi{K^{\mu}K^{\nu}
\over k_{0}^{2}}.\ee
The exact propagator (\ref{exact}) reduces in this gauge to
 \be
 D^{\prime\mu\nu}=-{A^{\mu\nu}\over
K^{2}-\Pi_{T}} +{K^{2}\over h_{0}^{2}}\left(\matrix{ 0&&0\cr
0&&\hat{k}^{m}\hat{k}^{n}\cr} \right)-\xi{H^{\mu}H^{\nu}\over
h_{0}^{2}},\label{tag}\ee 
where $h_{0}$ is the time component of $H^{\mu}$:
\be
 h_{0}=k_{0}\sqrt{K^{2}-\Pi_{L}}+\sigma k\sqrt{\Pi_{D}}.\ee
At momenta for which $h_{0}\to 0$, $H^{\mu}\to (0,\hat{k}h)$ and
$h\to \sigma\sqrt{\Pi_{D}}K^{2}/k_{0}$
so that the tensor structure of the residue is given by:
\be h_{0}^{2}\;D^{\prime\mu\nu}\to K^{2}[1-\xi\Pi_{D} {K^{2}\over k_{0}^{2}}]
\left(\matrix{ 0&&0\cr
0&&\hat{k}^{m}\hat{k}^{n}\cr} \right)\ee
As before $\Pi_{D}$ modifies the numerical value of the residue but not the spin
structure.

\section{Proof for General Linear Gauges}

\subsection{Gauge fixing condition}

The investigation pertains to any linear gauge-fixing condtion that is linear
and that does not break rotational invariance in the plasma rest frame. 
With a gauge-fixing term
${\cal L}_{\rm gf}=
-({\cal F}_{\mu}A^{\mu})^{2}/2\xi$ the quadratic part of the momentum-space
action is
 \be
{1\over 2}\int {d^{4}K\over (2\pi)^{4}}
\;A_{\mu}(K)\Gamma^{\mu\nu}_{\rm free}A_{\nu}(-K),\ee
where
\be
\Gamma^{\mu\nu}_{\rm free}=
-K^{2}g^{\mu\nu}+K^{\mu}K^{\nu}-{F^{\mu}F^{\nu}\over
\xi}.\label{freeact}\ee
  The most general $F^{\mu}$ that does not spoil the rotational invariance is a linear combination of
two vectors:  \be F^{\mu}=f_{1}K^{\mu}+f_{2}\tilde{K}^{\mu}\label{F},\ee
where $f_{1}$ and $f_{2}$ are functions of $k_{0}$ and $k$.
In the plasma rest frame $K^{\mu}=(k_{0},\vec{k})$ and $\tilde{K}^{\mu}=(k,k_{0}\hat{k})$.
It will be helpful to also introduce the vector
\be \tilde{F}^{\mu}=f_{1}\tilde{K}^{\mu}+f_{2}K^{\mu},\label{Ftilde}\ee
with the property
\be F\cdot\tilde{F}=0.\ee
For  the covariant, Coulomb, and temporal gauges the vectors (\ref{F}) and
(\ref{Ftilde}) are as follows:
\be\begin{array}{lcccc}
 															& F^{\mu}  &\tilde{F}^{\mu}  & f_{1}  & f_{2} \\
{\rm Covariant} & (k_{0},\vec{k}) & (k,k_{0}\hat{k})  & 1  & 0 \\
{\rm Coulomb}  & (0,\vec{k})   & (k,0)  & -k^{2}/K^{2} & k_{0}k/K^{2}  \\
{\rm Temporal} & (1,0)   & (0,\hat{k})  & k_{0}/K^{2}  & -k/K^{2}  
\end{array}\ee

\noindent The free propagator is the inverse of 
(\ref{freeact}):
\be 
D^{\mu\nu}_{\rm free}=-{A^{\mu\nu}\over K^{2}}
+{\tilde{F}^{\mu}\tilde{F}^{\nu}\over (f_{1}K^{2})^{2}}
-\xi{K^{\mu}K^{\nu}\over (f_{1}K^{2})^{2}}.\label{freeprop}\ee

\subsection{Exact gluon propagator}

For the full gluon propagator it is necessary  to add the self-energy to (\ref{freeact}) to obtain
the quadratic part of the full effective action:  
\be
\Gamma^{\mu\nu}=
-K^{2}g^{\mu\nu}+K^{\mu}K^{\nu}+\Pi^{\mu\nu}-{F^{\mu}F^{\nu}\over
\xi}.\label{act}\ee
The full gluon propagator, denoted $D^{\prime}_{\nu\alpha}$ is the inverse
\be
\Gamma^{\mu\nu}D^{\prime}_{\nu\alpha}=\delta^{\mu}_{\;\;\alpha}.\label{inv}\ee
The full propagator must satisfy the nonabelian Slavnov-Taylor identity
\be 
F^{\nu}F^{\alpha}D^{\prime}_{\nu\alpha}=-\xi,\label{STa}\ee
which is the generalization of (\ref{1.2}) to general linear gauges.
Appendix C shows that this identity must hold even at finite temperature.
To determine how (\ref{STa}) directly constrains  the proper self-energy,
it is useful to define a four-vector $G_{\nu}$ by
\be
 G_{\nu}\equiv D^{\prime}_{\nu\alpha}F^{\alpha}.\label{G}\ee
Then (\ref{inv}) implies that
\be
\Gamma^{\mu\nu}G_{\nu}=F^{\mu},\label{G2}\ee
and (\ref{STa}) implies
\be
F^{\nu}G_{\nu}=-\xi.\label{STb}\ee
Consequently (\ref{G2}) can be written
\be
\Bigl(\Gamma^{\mu\nu}+{F^{\mu}F^{\nu}\over \xi}\Bigr)G_{\nu}=0.\ee
The tensor in parentheses will be denoted by a check:
\begin{eqnarray}
\check{\Gamma}^{\mu\nu}&&\equiv \Gamma^{\mu\nu}+F^{\mu}F^{\nu}/\xi\cr\cr
&&=-K^{2}g^{\mu\nu}+K^{\mu}K^{\nu}+\Pi^{\mu\nu}.\end{eqnarray} 
In terms of the tensor decomposition (\ref{pi})
the explicit form  is
\begin{eqnarray} \check{\Gamma}^{\mu}_{\;\;\nu}&&=(\Pi_{T}-K^{2})A^{\mu}_{\;\;\nu}
+(\Pi_{L}-K^{2})B^{\mu}_{\;\;\nu}\cr\cr
&&+\Pi_{C}C^{\mu}_{\;\;\nu}+\Pi_{D}D^{\mu}_{\;\;\nu}.\label{act1}
\end{eqnarray}
Although $\check{\Gamma}$ has the gauge-fixing term removed it is still gauge-dependent because
of the self-energy functions.
It must satisfy
\be
\check{\Gamma}^{\mu}_{\;\nu}G^{\nu}=0,\ee
where $G^{\nu}$ is an unknown four-vector satisfying (\ref{STb}).
The first step in implementing this is to  requires that 
 $\check{\Gamma}$ have  determinant zero.
Since $\check{\Gamma}$ is a 4$\times$4 matrix, it must have four eigenvectors. Two of the
eigenvectors are  the two transverse polarization vectors $\epsilon^{\nu}$
which satisfy $K\cdot \epsilon=0$ and $ u\cdot\epsilon=0$. The eigenvalue for
each of these is the same:
$\check{\Gamma}^{\mu}_{\;\nu}\epsilon^{\nu}=(\Pi_{T}-K^{2})\epsilon^{\mu}$. The remaining
two eigenvectors   are each linear combinations of $K^{\nu}$ and
$\tilde{K}^{\nu}$. The properties of the basis tensors $A, B, C, D$ give
\begin{eqnarray}\check{\Gamma}^{\mu}_{\;\nu}K^{\nu}&&=\Pi_{C}\tilde{K}^{\mu}
+\Pi_{D}K^{\mu}\cr
\check{\Gamma}^{\mu}_{\;\nu}\tilde{K}^{\nu}&&=(\Pi_{L}-K^{2})\tilde{K}^{\mu}
-\Pi_{C}K^{\mu},\end{eqnarray}
which yields  a quadratic equation for the remaining two eigenvalues. 
The product of all four eigenvalues is  
\be {\rm Det}\bigl(\check{\Gamma}^{\mu}_{\;\nu}\bigr)=(\Pi_{T}-K^{2})^{2}
\Bigl[(\Pi_{L}-K^{2})\Pi_{D}+(\Pi_{C})^{2}\Bigr].\ee
The  Slavnov-Taylor identity forces the quantity in square brackets  to
vanish for all $K^{\mu}$:
\be
(\Pi_{L}-K^{2})\Pi_{D}+(\Pi_{C})^{2}=0.\ee
It seems most natural to solve this for $\Pi_{C}$:
\be
\Pi_{C}=\sigma\sqrt{(K^{2}-\Pi_{L})\Pi_{D}},\label{sig}\ee
where $\sigma=\pm1$.  When this condition is subsituted into (\ref{act1}), the particular
linear combination of the tensors $B, C, D$ can be written in terms of a single four-vector
$\tilde{H}^{\mu}$ as follows:
\begin{eqnarray}
\check{\Gamma}^{\mu\nu}&&=(\Pi_{T}-K^{2})A^{\mu\nu}+{\tilde{H}^{\mu}\tilde{H}^{\nu}
\over K^{2}}\cr\cr   
\tilde{H}^{\nu}&&\equiv 
\tilde{K}^{\nu}\sqrt{K^{2}-\Pi_{L}}+\sigma K^{\nu}\sqrt{\Pi_{D}}.
\end{eqnarray}
The null vector of $\check{\Gamma}^{\mu}_{\;\;\nu}$ is the vector 
\be
H^{\nu}=K^{\nu}\sqrt{K^{2}-\Pi_{L}}+\sigma\tilde{K}^{\nu}\sqrt{\Pi_{D}}.\ee It
satisfies $\check{\Gamma}^{\mu}_{\;\nu}H^{\nu}=0$ because 
$\tilde{H}\cdot H=0$.
(The null vector $G^{\nu}$ normalized by (\ref{STb}) is
$G^{\nu}=-\xi H^{\nu}/F\cdot H$.)

Now that the Slavnov-Taylor constraint has been satisfied, is is straightforward to 
 obtain the full propagator. First add back the gauge-fixing term  to
$\check{\Gamma}$: 
\be
\Gamma^{\mu\nu}=(\Pi_{T}-K^{2})A^{\mu\nu}+{\tilde{H}^{\mu}\tilde{H}^{\nu}
\over K^{2}}-{F^{\mu}F^{\nu}\over \xi}.\ee
It is easy to invert this using 
$F\cdot\tilde{F}= H\cdot \tilde{H}=0$ and $\tilde{F}\cdot\tilde{H}=-F\cdot H$ 
to obtain the full propagator: 
\be 
D^{\prime\mu\nu}=-{A^{\mu\nu}\over K^{2}-\Pi_{T}}
+K^{2}{\tilde{F}^{\mu}\tilde{F}^{\nu}\over(F\cdot H)^{2}}
-\xi{H^{\mu}H^{\nu}\over
(F\cdot H)^{2}}.\label{exact}\ee
Naturally it satisfies the Slavnov-Taylor identity (\ref{STa}). Sec II has displayed examples of
this result in covariant, Coulomb, and temporal gauges.

\section{Discussion}

\subsection{Gauge invariance of poles}

The tensor structure of the gluon propagator (\ref{exact}) is explicitly dependent
upon $\xi$ and upon the gauge-fixing parameters $f_{1}$ and $f_{2}$ contained in
$F^{\mu}$ and $\tilde{F}^{\mu}$. Moreover the self-energy $\Pi^{\mu\nu}$ is dependent upon these
same gauge parameters since they occur in the free propagator (\ref{freeprop}). Thus
all four  functions $\Pi_{T}$, $\Pi_{L}$, $\Pi_{C}$, and
$\Pi_{D}$ are gauge-dependent.

Kobes, Kunstatter, and Rebhan proved a remarkable theorem \cite{KKR}. If the exact propagator is
expressed as a linear combination of the four basis tensors $A^{\mu\nu}, B^{\mu\nu}, C^{\mu\nu},
D^{\mu\nu}$ then the location of the poles in the $A^{\mu\nu}$ and $B^{\mu\nu}$ part of the
propagator are  gauge-fixing independent. 
The transverse projection of (\ref{exact})  is
\be A_{\mu\nu}D^{\prime\mu\nu}=-{2\over K^{2}-\Pi_{T}}.\ee
The complex momentum that satisfy  $K^{2}=\Pi_{T}$ are gauge invariant by the KKR theorem.  
The $B_{\mu\nu}$ projection of (\ref{exact}) is
\be B_{\mu\nu}D^{\prime\mu\nu}=(\xi\Pi_{D}-f_{1}^{2}K^{2}){K^{2}\over (F\cdot H)^{2}}.\ee
The denominator is
\be
F\cdot H=K^{2}\Bigl[f_{1}\sqrt{K^{2}-\Pi_{L}}-\sigma f_{2}\sqrt{\Pi_{D}}\Bigr].\ee
This vanishes at complex $K$ that satisfy
\be
K^{2}=\Pi_{L}+\bigl({f_{2}\over f_{1}}\bigr)^{2}\Pi_{D}.\label{pole}\ee
Using (\ref{sig}) this can also be written in terms of $\Pi_{C}$ as 
$K^{2}=\Pi_{L}+\Pi_{C}(f_{2}/f_{1})$.
Despite the explicit gauge dependendent coefficients $f_{1}$ and $f_{2}$, the theorem of Kobes,
Kunstatter, and Rebhan \cite{KKR} shows that  the solution of this complex dispersion relation is
gauge invariant. At momenta which satisfy (\ref{pole}),
$H^{\mu}\to \tilde{F}^{\mu}\sqrt{K^{2}-\Pi_{L}}/f_{1}$.
Therefore the tensor structure of the propagator at the pole is simple:
\be
(F\cdot H)^{2}D^{\prime\mu\nu}\Big|_{\rm pole}\to\bigl[K^{2}-{\xi\over
f_{1}^{2}}(K^{2}-\Pi_{L})\bigr] \tilde{F}^{\mu}\tilde{F}^{\nu}.\ee

\subsection{Zero Three-Momentum}

The pole in the transverse sector occurs at $K^{2}=\Pi_{T}$. The pole in the longitudinal sector
occurs at (\ref{pole}). It is easy to show that at
$\vec{k}=0$ these poles coincide and that they have the same residue.  This comes about because the
proper self-energy $\Pi^{\mu\nu}(k_{0},\vec{k})$ has the property that when $|\vec{k}|=0$, it does
not depend upon $\hat{k}$. The proof of this is quite simple. The self-energy 
is computed by integrations over fermion and boson propagators. Let $P^{\alpha}$ denote a generic
loop momenta to be integrated  and $K^{\alpha}$ denote the external four-momentum of the gluon.
The fermion and boson propagators to be integrated over have arguments of the form
$P^{\alpha}+c_{j}K^{\alpha}$, where $c_{j}$ depends upon where the propagator occurs in the diagram.
At $|\vec{k}|=0$, the arguments of the propagators are $(p_{0}+c_{j}k_{0},\vec{p})$ and all
information about the direction $\hat{k}$ disappears. 

To apply this result, begin with the general self-energy
 \be \Pi^{\mu\nu}(k_{0},\vec{k})=\Pi_{T}A^{\mu\nu}+\Pi_{L}B^{\mu\nu}+\Pi_{C}C^{\mu\nu}
+\Pi_{D}D^{\mu\nu}.\ee
According to  Appendix A the $|\vec{k}|=0$ limit of the four basis tensors gives
for the spatial components of the self-energy
\begin{eqnarray}
\Pi^{mn}(k_{0},0)=&
(-\delta^{mn}+\hat{k}^{m}\hat{k}^{n})\Pi_{T}(k_{0},0)\cr
&-\hat{k}^{m}\hat{k}^{n}
\Pi_{L}(k_{0},0). \end{eqnarray}
Since this cannot depend upon $\hat{k}$
\be
\Pi_{T}(k_{0},0)=\Pi_{L}(k_{0},0).\ee
Similarly  
$\Pi^{0n}(k_{0},0)=\hat{k}^{n}\Pi_{C}(k_{0},0)$
and since this cannot depend on $\hat{k}$ it is necessary that 
$\Pi_{C}(k_{0},0)=0.$
Because of  relation (\ref{sig}) this automatically means that
$\Pi_{D}(k_{0},0)=0$.
Thus the self-energy tensor is
\be
\Pi^{\mu\nu}(k_{0},0)=\left(\matrix{0 && 0\cr
0&& -\delta^{mn}\Pi_{T}(k_{0},0)}\right).\ee
The full gluon propagator is quite simple in this limit. For example, in covariant gauge the
propagator is 
\be
D^{\mu\nu}(k_{0},0)=\left(\matrix{-\xi/k_{0}^{2} &0\cr
0& \delta^{mn}/[k_{0}^{2}-\Pi_{T}(k_{0},0)]}\right).\ee
It is also simple in the temporal gauge. In the Coulomb gauge the limit $|\vec{k}|\to 0$ does not
exist for either the free or the full propagator.

\acknowledgments

This work was supported in part by National Science Foundation grants
PHY-9213734 and PHY-9630149.

\appendix

\section{Tensor Basis}

There is a standard set of four tensors \cite{GPY,KK,rev} constructed out of
$g^{\mu\nu}, K^{\mu},$ and $\tilde{K}^{\mu}=(K\cdot u K^{\mu}-K^{2}u^{\mu})/k$ that will be used
throughout:  \begin{eqnarray}A^{\mu\nu}&&=g^{\mu\nu}-B^{\mu\nu}-D^{\mu\nu}\cr
B^{\mu\nu}&&=-{\tilde{K}^{\mu}\tilde{K}^{\nu}\over K^{2}}\cr
C^{\mu\nu}&&={K^{\mu}\tilde{K}^{\nu}+\tilde{K}^{\mu}K^{\nu}\over K^{2}}\cr
D^{\mu\nu}&&={K^{\mu}K^{\nu}\over K^{2}}\label{tensors}\end{eqnarray} 
The squares of these tensors are
\be A^{2}=A,\; B^{2}=B,\; C^{2}=-B-D,\; D^{2}=D\ee
Note that requiring $A, B$, and $D$ to be idempotent fixes their sign and
magnitude, but there is no such universal choice for $C$. The twelve mixed
products are all simple: \begin{eqnarray} &&AB=AC=AD=BD=0\cr
&&BA=CA=DA=DB=0\cr
&&(BC)^{\mu\nu}=(CD)^{\mu\nu}=\tilde{K}^{\mu}K^{\nu}/K^{2}\cr
&&(CB)^{\mu\nu}=(DC)^{\mu\nu}=K^{\mu}\tilde{K}^{\nu}/K^{2}\end{eqnarray}

It is often convenient to work in the rest frame of the plasma, where
$u^{\mu}=(1,0,0,0)$ and the vector $\tilde{K}^{\mu}$ takes the simple form
$(k,\hat{k}k_{0})$. In this frame the four basis tensors have the following values:
 \begin{eqnarray}
&&A^{\mu\nu}\Big|_{\rm rest}=\left(\matrix{ 0&& 0\cr 0 &&
-\delta^{mn}+\hat{k}^{m}\hat{k}^{n}\cr}\right) \cr\cr\cr
&&B^{\mu\nu}\Big|_{\rm rest}={1\over
K^{2}}\left(\matrix{-k^{2} && -k_{0}k^{n}\cr -k^{m}k_{0} &&
-k_{0}^{2}\hat{k}^{m}\hat{k}^{n}\cr}\right)\cr\cr\cr
&&C^{\mu\nu}\Big|_{\rm rest}
={1\over K^{2}}\left(\matrix{2k_{0}k &&
(k_{0}^{2}+k^{2})\hat{k}^{n}\cr
\hat{k}^{m}(k_{0}^{2}+k^{2}) && 2k_{0}k\hat{k}^{m}\hat{k}^{n}\cr}\right)\cr\cr\cr
&&D^{\mu\nu}\Big|_{\rm rest}
={1\over K^{2}}\left(\matrix{k_{0}^{2} && k_{0}k^{n}\cr
k^{m}k_{0} && k^{m}k^{n}\cr}\right)\end{eqnarray}

\section{Hard thermal loop approximation}

Although the structure of the gluon propagator in the hard thermal loop approximation is well known
in standard gauges, it is useful to obtain it in a general linear gauge using the method employed
in Sec III. Braaten and Pisarski showed that  the gluon propagator in the hard thermal loop
approximation obeys  abelian Ward identities \cite{BP1,BP2,F&T,act}.
In a    general linear gauge the abelian Ward identity is
\be
D^{\prime \rm htl}_{\nu\alpha}F^{\alpha}= -{\xi\over F\cdot K}K_{\nu}\label{abel},\ee
as shown in Appendix C. 
 This constraint  can be rewritten
by multiplying (\ref{abel}) by
 $\Gamma^{\mu\nu}_{\rm htl}$ to obtain 
\be
F^{\mu}=-{\xi\over F\cdot K}\,\Gamma^{\mu\nu}_{\rm htl}K_{\nu}.\ee
This is equivalent to
\be
\Bigl(\Gamma^{\mu\nu}_{\rm htl}+{F^{\mu}F^{\nu}\over\xi}\Bigr)K_{\nu}=0.\label{3aa}\ee
As before, denote the tensor in parenthesis by a check: 
$\check{\Gamma}^{\mu\nu}_{\rm htl}\equiv \Gamma^{\mu\nu}_{\rm htl}+F^{\mu}F^{\nu}/\xi$.
In terms of the tensor decomposition (\ref{pi})
\begin{eqnarray}
\check{\Gamma}^{\mu\;\rm htl}_{\;\;\nu}&&=(\Pi_{T}^{\rm htl}-K^{2})A^{\mu}_{\;\;\nu}
+(\Pi_{L}^{\rm htl}-K^{2})B^{\mu}_{\;\;\nu}\cr\cr
&&+\Pi_{C}^{\rm htl}C^{\mu}_{\;\;\nu}+\Pi_{D}^{\rm htl}D^{\mu}_{\;\;\nu}.
\label{mat}\end{eqnarray}
Condition (\ref{3aa}) requires $\check{\Gamma}^{\mu\;\rm htl}_{\;\;\nu}K^{\nu}=0$.
From the tensor definitions in Appendix A, 
  $A^{\mu}_{\;\;\nu}K^{\nu}=0$ and $B^{\mu}_{\;\;\nu}K^{\nu}=0$
so that $\check{\Gamma}^{\mu\;\rm htl}_{\;\nu
}K^{\nu}=\Pi_{C}^{\rm htl}\tilde{K}^{\mu} +\Pi_{D}^{\rm htl}K^{\mu}$.
Because the vectors $\tilde{K}^{\mu}$ and $K^{\mu}$ are linearly independent, the only solution is
$\Pi^{\rm htl}_{C}=0, \Pi^{\rm htl}_{D}=0$.
To obtain the  propagator in the hard thermal loop approximation, we must add the gauge-fixing term
to (\ref{mat}): 
 \be
\Gamma^{\mu\nu}_{\rm
htl}=(\Pi_{T}^{\rm htl}-K^{2})A^{\mu\nu}+(\Pi_{L}^{\rm htl}-K^{2})B^{\mu\nu}
 -{F^{\mu}F^{\nu}\over \xi}.\ee 
The inverse of this is the hard thermal loop propagator.
Using $F\cdot \tilde{F}=K\cdot\tilde{K}=0$ and $F\cdot K=-\tilde{F}\cdot\tilde{K}
=f_{1}K^{2}$ gives the result
\be
D^{\prime\mu\nu}_{\rm htl}={-A^{\mu\nu}\over K^{2}-\Pi_{T}^{\rm htl}}
+{\tilde{F}^{\mu}\tilde{F}^{\nu}\over (f_{1})^{2}K^{2}(K^{2}-\Pi_{L}^{\rm htl})}
- {\xi K^{\mu}K^{\nu}\over (f_{1}K^{2})^{2}}.\ee

\section{Slavnov-Taylor Identity in General Linear Gauges}

This Appendix will derive the Slavnov-Taylor identity for the gluon
propagator at finite temperature.
The generating functional for the real-time Green functions of
finite-temperature QCD is \cite{rev}
\be 
Z(J)=\int D\Omega\;\exp(iS_{J}),\ee
where
\be
S_{J}=\int_{C}d^{4}x\bigl({\cal L}-{1\over 2\xi}({\cal F}_{\nu}A^{\nu})^{2}
+J_{\nu}A^{\nu}\bigr).\ee
The  label $C$ denotes a contour  in the complex
time plane that runs from an initial time $t_{0}$ and ends at $t_{0}-i\beta$
along a path that will be specified later. 
The gauge-fixing vector ${\cal F}_{\nu}$ determines the 
 Fadeev-Popov matrix
\be
{\cal M}(x,y)=-{\cal F}_{x\nu}D^{\nu}_{x}\delta^{4}(x-y)\ee
and the invariant integration measure
\be
D\Omega=D\psi D\overline{\psi}DA\; {\rm det}({\cal M}).\ee
The generating functional is invariant under a change of the integration
variables $A^{\mu}$. If the change of variables is a gauge transformation
$A^{\mu}\to A^{\mu}+D^{\mu}\Lambda$ then the 
 measure $D\Omega$ and the quark-gluon lagrangian ${\cal L}$ will not change.
The only change  comes from the gauge-fixing and source
terms. Thus
\be 
0=\int D\Omega\,\delta S_{J}
\;\exp(iS_{J})\label{dZ}\ee
where
\be 
\delta S_{J}=\int_{C}d^{4}x
\Bigl( 
-{({\cal F}_{\mu}A^{\mu}){\cal F}_{\nu}D^{\nu}\Lambda\over\xi}
+J_{\nu}D^{\nu}\Lambda\Bigr).\ee
It is convenient to choose $\Lambda$ as
\be \Lambda(x)=\int_{C} d^{4}y\; [{\cal F}_{\mu}D^{\mu}]^{-1}(x,y)\;\lambda(y)\ee
for an arbitrary function $\lambda(y)$ so that
\be
{\delta S_{J}\over\delta\lambda(y)}=-{1\over\xi}{\cal F}_{\nu}A^{\nu}(y)
+\int_{C}d^{4}x\;J_{\nu}(x)D^{\nu}_{x}[{\cal F}\cdot D]^{-1}(x,y)\ee
The invariance (\ref{dZ}) becomes
\be
{{\cal F}_{\nu}(y)\over Z(0)}{i\delta Z\over\delta J_{\nu}(y)}
=\int_{C}d^{4}x\;J_{\nu}(x)G^{\nu}(x,y)\label{dZ2}\ee
\be
G^{\nu}(x,y)\equiv
-\xi\int D\Omega D^{\nu}_{x}[{\cal F}\cdot D]^{-1}(x,y)
\exp(iS_{J})/Z(0).\label{defG}\ee
This has the property that
\be
{\cal F}_{\mu}(x)G^{\mu}(x,y)\Big|_{J=0}=-\xi\;\delta_{C}^{4}(x-y).\ee
Now apply $\delta/\delta J_{\mu}(x)$ to (\ref{dZ2})
\be
{{\cal F}_{\nu}(y)\over Z(0)}{i\delta^{2} Z\over\delta J_{\mu}(x)\delta J_{\nu}(y)}
=G^{\mu}(x,y)+\int_{C}d^{z}y J_{\alpha}(z){\delta G^{\alpha}(z,y)\over\delta J_{\mu}(x)}
\label{der2}\ee
and set $J=0$:
\be
{\cal F}_{\nu}(y)D^{\prime\mu\nu}(x,y)=
G^{\mu}(x,y)\Big|_{J=0}.\label{Gmu}\ee
In QCD the covariant derivative in (\ref{defG}) depends on the
vector potential.  The Slavnov-Taylor identity follows
from applying ${\cal F}_{x\mu}$ to (\ref{Gmu}):
\be
{\cal F}_{\mu}(x){\cal F}_{\nu}(y)D^{\mu\nu}(x,y)=-\xi\;\delta^{4}_{C}(x-y).\label{coor}\ee

To convert this to momentum space it is necessary to know where
 the time components of $x$ and $y$ lie on the contour $C$ in the
complex time plane. For field theory in Minkowski space  the contour $C$ is taken
as the union of four parts \cite{rev} that depends on a parameter $\sigma$ with
$0\le\sigma\le \beta$: $C_{1}$ runs from
$-t_{0}$ to $t_{0}$, $C_{3}$ runs from $t_{0}$ to $t_{0}-i\sigma$,
$C_{2}$ runs from $t_{0}-i\sigma$ to $-t_{0}-i\sigma$, and $C_{4}$ runs from
$-t_{0}-i\sigma$ to $-t_{0}-i\beta$. As $t_{0}\to\infty$ the contributions of
$C_{3}$ and $C_{4}$ can be neglected. The physical
field $A^{\mu}_{1}(x)\equiv A^{\mu}(x)$ and the
auxiliary field $A^{\mu}_{2}(x)\equiv A^{\mu}(t-i\sigma,\vec{x})$ lie on $C_{1}$ and
$C_{2}$, respectively. 
Since the time components of $x$ and $y$  may be taken along
$C_{1}$ or $C_{2}$ independently, the propagator becomes a $2\times 2$ matrix
in the internal space. The contour-ordered propagator contains four time-ordered parts.
The explicit form of (\ref{coor}) is
\be
{\cal F}_{\mu}(x){\cal F}_{\nu}(y)D^{\prime\mu\nu}_{ij}(x,y)=-\xi\;\delta^{4}(x-y)(\tau_{3})_{ij}
\ee
where now $x$ and $y$ have real time components. In momentum space this becomes
\be F_{\mu}F_{\nu}D^{\prime\mu\nu}_{ij}(K)=-\xi(\tau_{3})_{ij}.
\label{ST2}\ee
For covariant gauges $F^{\mu}=K^{\mu}$, for Coulomb gauges $F^{\mu}
=(0,\vec{k})$, and for temporal gauge $F^{\mu}=(1,0,0,0)$.

\underbar{Feynman Basis:} The $2\times 2$ matrix structure of the time-ordered
thermal propagator may be diagonalized in terms of functions which have
Feynman-like analytic structure (analytic in the upper half-plane of
complex $(k_{0})^{2}$) as follows \cite{rev,EKW}: 
\be D^{\prime\mu\nu}_{ij}(K)=U_{i\ell}
\left(\matrix{D_{F}^{\prime\mu\nu}(K)&&0\cr
 0&&-D_{F}^{\prime\mu\nu\ast}(K)\cr}\right)_{\ell m}
V_{mj},\label{Feyn}\ee
where $U$ and $V$ are nonsingular matrices that satisfy $U\tau_{3}V=\tau_{3}$. 
 The matrix equation (\ref{ST2}) therefore reduces
to the single condition
\be F_{\mu}F_{\nu}D^{\prime\mu\nu}_{F}(K)=-\xi.\ee
Condition (\ref{ST2}) imposes no addtional constraing on the  off-diagonal parts of the propagator
 because $D^{\prime\mu\nu}_{12}\propto {\rm Im} D^{\prime\mu\nu}_{11}$.
Therefore $F_{\mu}F_{\nu}D^{\prime\mu\nu}_{12}\propto
 F_{\mu}F_{\nu}{\rm Im} D^{\prime\mu\nu}_{11}\propto {\rm Im}\xi=0$.

\underbar{Retarded Basis:} The time-ordered propagator may also be diagonalized
in terms of retarded propagators (analytic in the upper half-plane of complex
$k_{0}$) and advanced propagators (analytic in the lower half-plane of complex
$k_{0}$) \cite{Aurenche}:
\be D^{\prime\mu\nu}_{ij}(K)=U_{i\ell}
\left(\matrix{D_{R}^{\prime\mu\nu}(K)&&0\cr
 0&&-D_{A}^{\prime\mu\nu}(K)\cr}\right)_{\ell m}
V_{mj}\ee
where  $U$ and $V$ are different from those in  (\ref{Feyn}) but still satisfy
 $U\tau_{3}V=\tau_{3}$.
As before the matrix equation (\ref{ST2})  reduces
to a single condition on the retarded function
\be F_{\mu}F_{\nu}D^{\prime\mu\nu}_{R}(K)=-\xi.\ee

\underbar{Abelian Ward Identity:} For an abelian theory the covariant derviative
in (\ref{defG}) is independent of the vector potential so that (\ref{Gmu}) becomes
\be
{\cal F}_{\nu}(y)D^{\prime\mu\nu}(x,y)=-\xi{\partial\over\partial x_{\nu}}
\bigl[{\cal F}\cdot \partial\bigr]^{-1}(x,y).\ee
In momentum space this is
\be
F_{\nu}D^{\prime\mu\nu}_{ij}(K)=-\xi\; {K^{\mu}\over F\cdot K}\;(\tau_{3})_{ij},\ee
which may be diagonalized in either the Feynman or the retarded basis.

\end{document}